\documentclass[12pt]{article}
\usepackage{graphicx}

\newcommand\pubnumber{}
\newcommand\pubdate{\today}

\def\urbino{Universit\`a di Urbino and INFN Sezione Firenze\\
Via S. Chiara 27, I--61029 Urbino, Italy}

\def\support{\footnote{On behalf of the NA48/2 collaboration: 
Cambridge, CERN, Chicago, Dubna, Edinburgh, Ferrara, Firenze,
Mainz, Northwestern, Perugia, Pisa, Saclay, Siegen, Torino, Vienna.}}

\def\Title#1{\begin{center} {\Large #1 } \end{center}}
\def\Author#1{\begin{center}{ \sc #1} \end{center}}
\def\Address#1{\begin{center}{ \it #1} \end{center}}

\newcommand\pubblock{\rightline{\begin{tabular}{l} \pubnumber\\
         \pubdate  \end{tabular}}}
\newenvironment{Abstract}{\begin{quotation}  }{\end{quotation}}
\newenvironment{Presented}{\begin{quotation} \begin{center}
             PROCEEDINGS OF\end{center}\bigskip
      \begin{center}\begin{large}}{\end{large}\end{center} \end{quotation}}

\newcommand{\vus}{$\vert V_{us} \vert$~}

\newcommand{\kche}{$\rm K^{\pm}_{e3}$}
\newcommand{\kchm}{$\rm K^{\pm}_{\mu3}$}

\begin{document}

\begin{titlepage}
\pubblock

\vfill
\Title{$\rm K^{\pm}_{\mu3}$ Form Factors Measurement at NA48/2}
\vfill
\Author{ Michele Veltri\support}
\Address{\urbino}
\vfill
\begin{Abstract}
We report here a measurement of form factors of $K^{\pm}_{\mu 3}$
decay by the NA48/2 experiment at CERN. 
Using a sample of 3.4$\times10^6$ events we provide preliminary
form factor values according to various parametrizations. 
The slope of the scalar form factor is in agreement with other 
measurements and theory predictions.
\end{Abstract}
\vfill
\begin{Presented}
CKM2010, the $6^{th}$ International Workshop on the \\
CKM Unitarity Triangle, University of Warwick, UK, \\
6--10 September 2010
\end{Presented}
\vfill
\end{titlepage}
\def\thefootnote{\fnsymbol{footnote}}
\setcounter{footnote}{0}

\section{Introduction}
In recent years semileptonic kaon decays ($\rm K_{\ell 3}, \ell=e, \mu$)
have attracted renewed interest  \cite{WG1-2010}.
These decays provide the most accurate and theoretically cleanest way 
to measure \vus and can give stringent constraints on new physics 
scenarios by testing for possible violations of CKM unitarity and 
lepton universality.
The hadronic matrix element of these decays is described by two
dimensionless form factors $f_{\pm}(t)$ where $t=(p_{K}-p_{\pi})^2$
is the four--momentum squared transferred to the lepton system.
These form factors are one of the input (through the phase space 
integral)  needed to determine $\vert V_{us} \vert$.
In the matrix element $f_{-}$ is multiplied by the lepton mass
and therefore its contribution can be neglected in $\rm K_{e3}$ 
decays.
$\rm K_{\mu3}$ decays instead are usually described in terms of 
$f_{+}$ and the scalar form factor $f_{0}$ defined as:
\begin{equation}
 f_{0}(t) = f_{+}(t) + t/(m^{2}_{K} - m^{2}_{\pi}) f_{-}(t), \nonumber
\end{equation}
$f_{+}$ and $f_{ 0}$ are related to the vector ($1^{-}$) and scalar
($0^{+}$) exchange to the lepton system, respectively.
By construction $f_{0}(0)=f_{+}(0)$ and since $f_{+}(0)$ is not 
directly measurable it is customary to factor out $f_+^{K^0 \pi^-}(0)$
and normalize to this quantity all the form factors so that:
\begin{equation}
 \bar f_+(t) = \frac{f_+(t)}{f_+(0)},
 \ \bar f_0(t) = \frac{f_0(t)}{f_+(0)},
 \ \bar f_+(0) = \bar f_{0} (0) = 1. \nonumber
\label{eq:Normff}
\end{equation}
There exist many parametrizations of the $\rm K_{\ell 3}$ form factors
in the literature, a widely known and most used is the Taylor
expansion:
\begin{equation}
 \bar f_{+,0}(t) = 1 +
 \lambda'_{+,0} \frac{t}{m_{\pi^\pm}^2} +
 \frac{1}{2}\lambda''_{+,0} \left(\frac{t}{m_{\pi^\pm}^2}\right)^2, \nonumber 
\label{eq:Taylor}
\end{equation}
where $\lambda'_{+,0}$ and $\lambda''_{+,0}$ are the slope and
the curvature of the form factors, respectively.
The disadvantage of such kind of parametrization is related \cite{franzini}
to the strong correlations that arise between parameters.
These forbid the experimental determination of  $\lambda''_{0}$
experimentally, although, at least a quadratic expansion would be needed to 
correctly describe the form factors. 
This problem is avoided by parametrizations which, applaying 
physical constraints, reduce to one the number of parameters used. 
A typical example is the pole one:
\begin{equation}
 \bar f_{+,0}(t) = \frac{M_{V,S}^2}{M_{V,S}^2-t}, \nonumber
\label{eq:pole}
\end{equation}
where the dominance of a single resonance is assumed and the 
corresponding pole mass $M_{V,S}$ is the only free parameter.
More recently a parametrization based on dispersion techniques
has been proposed \cite{stern}:
\begin{eqnarray}
 \bar{f}_{+}(t) = \exp\Bigl{[}\frac{t}{m^{2}_{\pi}}(\Lambda_{+} 
                    + \mathrm{H(t)})\Bigr{]},~~~~~~~~~
 \bar{f}_{0}(t) = \exp\Bigl{[}\frac{t}{\Delta_{K\pi}}
                  (\mathrm{lnC}-\mathrm{G(t)})\Bigr{]}. \nonumber
\label{eq:dispersive}
\end{eqnarray}
The parameter C is the value of the scalar form factor at the 
Callan--Treiman point, $f_{0}(t_{CT})$, where
$t_{CT} = \Delta_{K\pi} = m_{K}^{2} - m_{\pi}^{2}$.
It can be used to test the existence of right handed quark 
currents coupled to the standard W boson.\\

\section{The NA48/2 experimental set--up}
\begin{figure}[htb]
\centering
\includegraphics[height=5.cm]{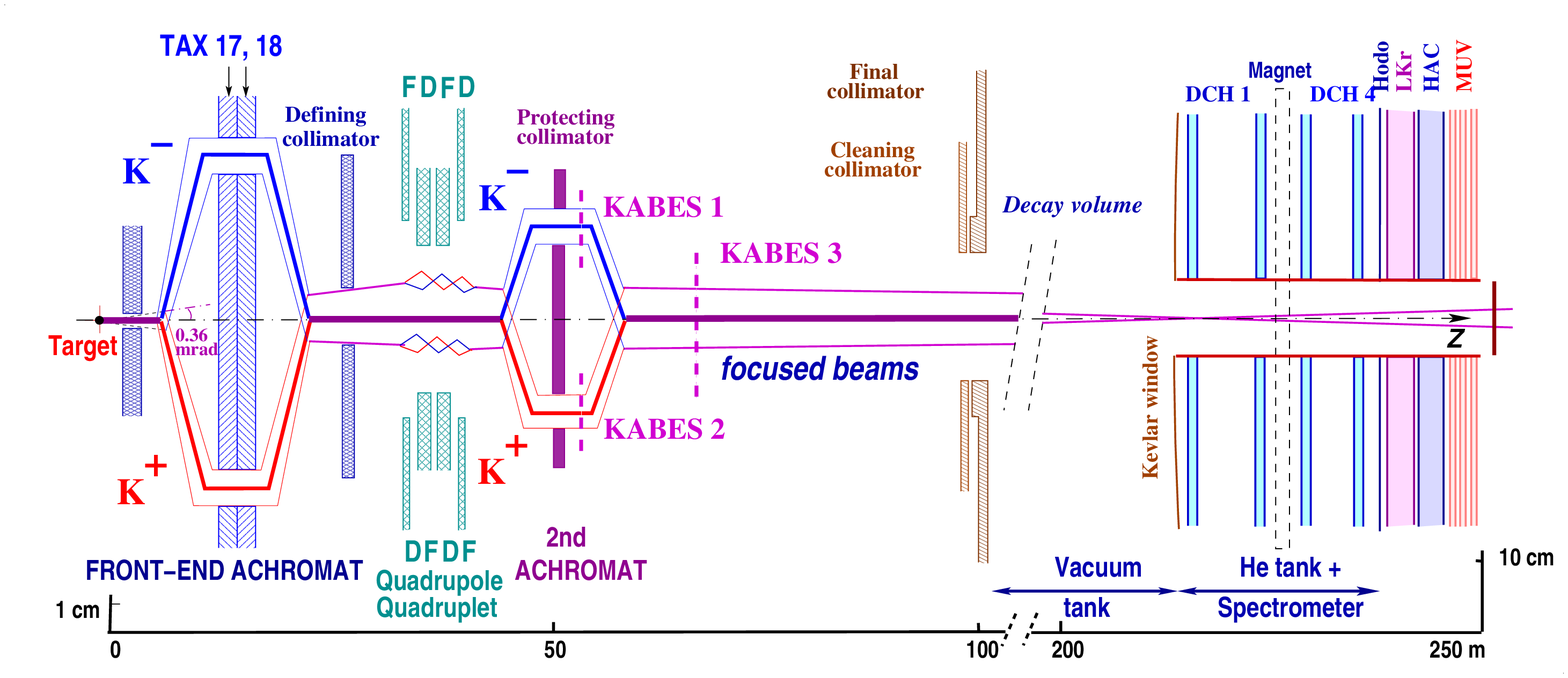}
\caption{Schematic side view of the NA48/2 beam line, decay volume and 
detectors.}
\label{fig:na48-2}
\end{figure}

The NA48/2 experiment at CERN/SPS was primarly designed to measure 
the CP violating asymmetry in $\rm K^{\pm} \to 3\pi$ decays. 
The layout of beams and detectors is shown on Fig.~\ref{fig:na48-2} 
Two simultaneous $\rm K^+$ and $\rm K^-$ beams were produced by 400 GeV
primary protons impinging on a beryllium target. A system of magnets
and collimators selected particles, with average momentum of 60~GeV,
of both positive and negative charge. At the entrance of the decay
volume, a 114~m long vacuum tank, the $\rm K^+$ flux was $\sim2.3\times10^6$
per pulse of 4.5~s duration and the $\rm K^+/ \rm K^-$ ratio was about 1.8.
The NA48 detector is described in detail elsewhere \cite{NA48detector},
the main component used in this analysis are:\\
-- a magnetic spectrometer (DCH), designed to measure the momentum of 
charged particles, it consisted of a magnet dipole located 
between two sets of drift chambers. The obtained momentum resolution was
 $\sigma(p)/p(\%) = 1 \oplus 0.044~p$ ($p$ in GeV/c);\\
-- a charged hodoscope (Hodo), made of two perpendicular segmented planes
of scintillators, it triggered the detector readout. The time resolution
was $\sim 150~$ps;\\
-- a liquid krypton electromagnetic calorimeter (LKr) of 27 radiation 
lengths and e\-ner\-gy resolution of 
$\sigma(E)/E(\%) = 3.2/\sqrt{E} \oplus 9.0/E \oplus 0.42$ ($E$ in GeV);\\
-- a muon system (MUV) consisting of three planes of alternating 
    horizontal and vertical scintillator strips, each plane was shielded 
    by a 80~cm thick iron wall. The inefficiency of the system was at
    the level of one per mill and the time resolution was below 1~ns.
The data used for this analysis were collected in 2004 during a dedicated
run with a special trigger setup, 
lower intensity 
and a reduced momentum spread.

\section{ \kchm~ event selection}
\kchm~events are selected by requiring a track in DCH and at least
 two clusters (photons) in  LKr that are consistent with a $\pi^0$
decay. The track has to be inside the geometrical acceptance of the 
detector, satisfy vertex and timing cuts and have $p >$ 10 GeV/c to
ensure proper efficiency of MUV system. In order to be identified 
as a muon the track has to be associated in space and time to a MUV 
hit and have $E/p < 0.2$, where $E$ is the energy deposited in the 
calorimeter and $p$ the track momentum. 
Finally a kinematical constraint is applied requiring the missing mass 
squared (in the $\mu$ hypothesis) to satisfy: $| m_{\nu} | <$ 10~MeV$^2$.
Background from $\rm K^{\pm} \to \pi^{\pm} \pi^0$ events with 
charged $\pi$ that decays in flight are suppressed by using a 
combined cut on the invariant mass $m_{\pi^{\pm} \pi^0}$ and the 
$\pi^0$ transverse momentum. 
This cut reduces to 0.6\% the contamination but causes a
loss of statistics of about 24\%. 
Another source of background is due to 
$\rm K^{\pm} \to \pi^{\pm} \pi^0 \pi^0$ events with $\pi$
decaying in flight and a $\pi^0$ not reconstructed, the estimated 
contamination amounts to about 0.1\% and no specific cut is applied. 
The selected \kchm~sample amounts to about 3.4$\times 10^6$ events.

\section{Fitting procedure and preliminary results}
\begin{figure}[h]
\centering
\vskip -1.0cm 
\includegraphics[height=11.cm]{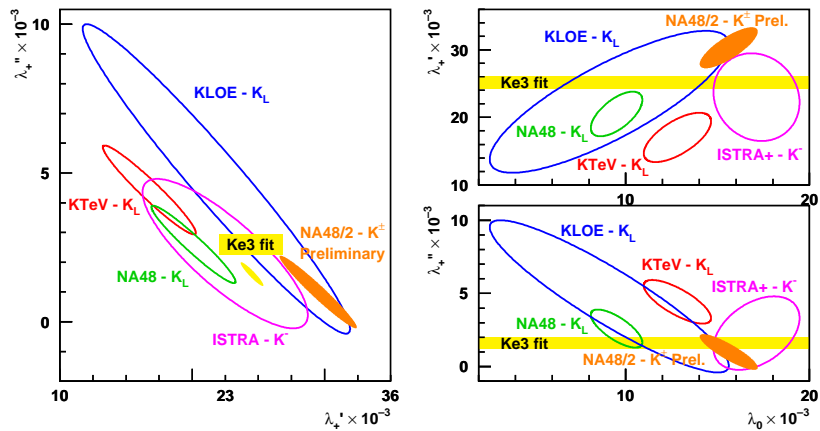}
\vskip -4.5cm 
\caption{Quadratic fit results for $\rm K_{\mu 3}$ ($\rm K_{L}$ for neutral  
and $\rm K^{\pm}$ for charged) decays. 
The ellipses are the 1 $\sigma$ contour plot. 
For comparison also the $\rm K_{e3}$ fit from FlaviaNet 
WG1 is shown.}
\label{fig:kmu3ff}
\end{figure}

To extract the form factors a fit is performed to the Dalitz
plot density. The Dalitz plot is subdivided into 
5$\times$5 MeV$^2$ cells, those crossed by the kinematical 
border are not used for the fit. 
The raw density must be corrected for acceptance and resolution,
residual background, and the distortions induced by radiative effects.
The results of the fit for quadratic, pole and dispersive 
parametrizations, are listed in Table~\ref{table:fit_results}.
\begin{table}[h]
\begin{center}
\begin{tabular}{cccc}
\hline
\rule{0mm}{5mm}Quadratic ($\times 10^{3}$) & 
  $\lambda^{'}_{+}$    & $\lambda^{''}_{+}$  & $\lambda_{0}$ \\
& 30.3$\pm$2.7$\pm$1.4  & 1.0$\pm$1.0$\pm$0.7 & 15.6$\pm$1.2$\pm$0.9 \\ \hline
\rule{0mm}{5mm}Pole (MeV/c$^2$) &
  $m_V$             &      $m_S$         &  \\
& 836$\pm7\pm9$   & 1210$\pm$25$\pm$10 &  \\ \hline
\rule{0mm}{5mm}Dispersive ($\times 10^{3}$) &
$\Lambda_{+}$          & $\ln C$                &  \\
& 28.5$\pm$0.6$\pm$0.7$\pm$0.5  & 188.8$\pm$7.1$\pm$3.7$\pm$5.0 &  \\ \hline
\end{tabular}
\caption{ NA48/2 preliminary form factors fit results for quadratic, 
pole and dispersive parametrizations. The first error is statistical, 
the second systematical. The theo\-re\-ti\-cal uncertainty \cite{stern} 
has been evaluated and added to dispersive results.}
\label{table:fit_results}
\end{center}
\end{table}
The comparison with the results of $\rm K_{\mu 3}$ quadratic fit as 
reported by recent experiments \cite{WG1-2010} is shown in 
Fig.~\ref{fig:kmu3ff}.
The $1 \sigma$ contour plots are displayed both for $\rm K^{0}_{\mu 3}$ 
decays (KLOE, KTeV and NA48) and charged $\rm K$ (ISTRA studied 
$\rm K^{-}_{\mu3}$ only), our high precision measurement is the 
first to use both $\rm K^+$ and $\rm K^-$ particles. 
We find a quadratic term in the expansion of the vector form 
factor compatible with zero and a slope of the scalar form factor 
larger with respect to NA48 case \cite{NA48kmu3ff} and in agreement 
with other measurements.
For this preliminary evaluation of systematic uncertainty
we have changed by small amounts the cuts defining the vertex quality 
and the geometrical acceptance, we applyed variations to the
values of pion and muon energies in the kaon cms, we increased  
$\pi \to \mu$ background and took into account the differences in the 
results of two independent analyses that are realized in parallel. 
\section{$\rm K^{\pm}_{e3}$ form factors and future perspectives at NA62}
Using the same data sample we are also investigating \kche~decays. Their 
selection is similar to that of \kchm, a track and a $\pi^0$ being 
required. The electron ID is achieved by demanding $0.95<E/p<1.05$,
this results in a \kche~sample of 4.2$\times 10^6$ events. 
Since these decays are described by only one form factor the problems 
related to the correlations between parameters are greatly reduced. 
Furthermore background issues are less critical given that to fake these 
decays $\pi^{\pm} \pi^0$ events with a $\pi^{\pm}$ having $E/p>0.95$ 
are needed. For these reasons, results of higher precision with respect
to \kchm~analysis are expected from this measurement.\\
The NA62 experiment, using the same beam line and detector of NA48/2, 
collected in 2007 data for the measurement of 
$R_K = \Gamma(\rm K_{e2})/\Gamma(\rm K_{\mu2})$ and made tests for the future 
$\rm K^{+} \to \pi^{+} \nu~ \overline \nu$ experiment. The data collected
contain also  $\rm K^{+}_{e3}/\rm K^{+}_{\mu 3}$ samples of 
$\simeq 40/20\times 10^6 $ events.
A special  $\rm K_L$ run was also taken, it provides 
$\rm K^{0}_{e 3}$ and $\rm K^{0}_{\mu 3}$ sample 
of about $4\times 10^6$ events. 
With this statistics, high precision measurements of the form factors for 
all $\rm K_{\ell 3}$ channels will be done by NA62, providing important 
inputs to further reduce the uncertainty on \vus.



\begin{thebibliography}{99}


\bibitem{WG1-2010}
  M.~Antonelli {\it et al.},
  arXiv:1005.2323 [hep-ph].

\bibitem{franzini}
P. Franzini, PoS KAON (2008) 002.

\bibitem{stern}
  V.~Bernard, M.~Oertel, E.~Passemar and J.~Stern,
  Phys.\ Lett.\  {\bf B638 } (2006)  480. and
  Phys.\ Rev.\  {\bf D80 } (2009)  034034.

\bibitem{NA48detector}
  V.~Fanti {\it et al.} [NA48 Collaboration],
  Nucl.\ Instrum.\ Meth.\  {\bf A574 } (2007) 433.
  
\bibitem{NA48kmu3ff}
  A.~Lai {\it et al.} [NA48 Collaboration],
  Phys.\ Lett.\  {\bf B647 } (2007) 341.

\end{thebibliography}
\end{document}